\documentclass[11pt,a4paper]{article}

\usepackage{graphicx}
\usepackage{amsmath}
\usepackage{ams}

\def\mineappendix{
        \setcounter{section}{1}
        \setcounter{subsection}{0}
        \def\thesection{\Alph{section}}
        \def\sectionap{\@startsection  {section}{1}{\z@}
                        {-3.5ex plus-1ex minus-.2ex} {0ex plus.2ex}
                        {\reset@font\Large\bf  Appendix:  \, }
                        }
        }
\makeatother
\def\Proclaim #1. #2\par{\bigbreak\noindent{\sc#1.\enspace}{\it#2}\par}

\font\Bbbfont=msbm10
\newfam\msbfam
\textfont\msbfam=\Bbbfont \textfont\msbfam=\Bbbfont

\def\al{\alpha}
\def\diag{\mbox{diag}}

\def\diag{\mbox{diag}}

\def\nn{\nonumber}

\def\nd{\noindent}

\usepackage{graphicx}
\usepackage{tabls}
\usepackage{amsmath}
\usepackage{amssymb}
\usepackage{bm}

\renewcommand\baselinestretch{1.2}

\title{The Transmission Property of the Discrete Heisenberg Ferromagnetic Spin Chain}
\author{
 Qing Ding\thanks{qding@fudan.edu.cn}~ and Wei Lin\\
Inst. of Math. and Key Lab. of Math. for Nonlinear
Sciences  \\
Fudan University, Shanghai 200433, P.R. China}
\date{}
\begin{document}
\maketitle

\begin{abstract}

We present a mechanism for displaying the transmission property of
the discrete Heisenberg ferromagnetic spin chain (DHF) via a
geometric approach. By the aid of a discrete nonlinear
Schr\"odinger-like equation which is the discrete gauge equivalent
to the DHF, we show that the determination of transmitting
coefficients in the transmission problem is always bistable. Thus a
definite algorithm and general stochastic algorithms are presented.
A new invariant periodic phenomenon of the non-transmitting behavior
for the DHF, with a large probability, is revealed by an adoption of
various stochastic algorithms.

\bigskip
\nd PACS numbers: 02.40.Ky; 05.45.Mt; 07.55.Db

%
\end{abstract}

\section*{\S1 Introduction}
There is current interest in displaying the properties of
one-dimensional magnetic models. The one-dimensional classical
continuum Heisenberg models with different magnetic interactions
have been settled as one of the interesting and attractive class of
nonlinear dynamical equations exhibiting the complete integrability
on many occasions (\cite{PN}-\cite{dq4}). However, though the
investigation of nonlinear spin chain systems is quite fascinating,
few has been known in the case of more realistic physical lattice
spin chains so far, especially, for the following discrete
(isotropic) Heisenberg ferromagnetic spin chain with nearest
neighbor exchange interaction (DHF),
\begin{eqnarray}\label{1}
\dot{\bf S}_n={\bf S}_n\times({\bf S}_{n+1}+{\bf S}_{n-1}),
\end{eqnarray}
where ${\bf S}_n =(s_n^1,s_n^2,s_n^3)\in {\bf
R}^3~~\hbox{with}~~(s^1_n)^2+(s^2_n)^2+(s^3_n)^2=1$ and the dot
stands for the time derivative.
In fact Eq.(\ref{1}) comes from the Hamiltonian formalism: ${\dot
{\bf S}_n}=\{{\bf S}_n,H\}$ with the Hamiltonian function
$H=\sum_{j=-\infty}^{+\infty}{\bf S}_j\cdot{\bf S}_{j+1}$, where
~$\cdot$~ denotes the inner product of vectors in ${\bf R}^3$, and
the Poisson bracket
$\{s_j^a,s_k^b\}=\delta_{jk}\sum_{c=1}^3\varepsilon_{abc}s_j^c,$
where $\delta_{jk}$ is Kronecker's symbol and $\varepsilon_{abc}$ is
the 3-dimensional totally antisymmetric symbol. Physically, one
would say the model (\ref{1}) describes a system of classical spins
and subjects to homogenous nearest-neighbor Heisenberg interaction.
The standard continuous limit procedure performed on Eq.(\ref{1})
leads to the integrable Heisenberg ferromagnetic model: ${\bf
S}_t={\bf S}\times {\bf S}_{xx}$, which is an important equation on
condensed matter physics (see, for example, \cite{MS}). Eq.(\ref{1})
is quite well known and there is little need to give here a detailed
enumeration. It should be also emphasized that Eq.(\ref{1}) is
widely believed to be not a completely integrable equation. To our
best knowledge, there is no effective method in study of the
dynamical behaviors of Eq.(\ref{1}) in literature.

On the other hand, the geometric concept of gauge equivalence
\cite{ZT,FT} between integrable equations, which provides a useful
tool in displaying solitonic dynamics, has been generalized to
nonintegrable case in \cite{dq5,dq6}. In \cite{dq6} it is shown that
the discrete nonlinear (nonintegrable) Schr\"odinger equation
(AL-DNLS) is the discrete gauge equivalent to a nonintegrable
discrete Heisenberg model and the transmission properties of the
AL-DNLS equation (\cite{DLS}-\cite{KRB}) are completely preserved
under the action of discrete gauge transformations. In this paper,
we would apply this geometric idea to transform the DHF (\ref{1})
into a discrete nonlinear Schr\"odinger-like equation and utilize
the nonlinear Schr\"odinger-like equation to study the transmission
problem of the DHF (\ref{1}) from mathematical point of view. We
hope that this study will be helpful to reveal deeper quantum
chaotic properties of the DHF (\ref{1}) and useful in physical
applications.

This article is organized as follows. In section 2 we deduce a
discrete nonlinear Sch\"odinger-type equation which is the discrete
gauge equivalent to the DHF in the category of the (discrete)
Yang-Mills theory. In section 3 we display the transmission property
for the DHF (\ref{1}) with the aid of the discrete nonlinear
Sch\"odinger-type equation and an approximate linear stability
analysis for the stationary version of the discrete nonlinear
Sch\"odinger-type equation is given for supporting the discussion of
the transmission expositions.  Finally, in section 4, we close the
paper with some conclusions and remarks.

\section*{\S2 Gauge equivalence}

Following the concept of discrete connection and associated discrete
curvature introduced in \cite{dq6}, we would express Eq.(\ref{1})
geometrically as a discrete nonlinear equation with prescribed
discrete curvature representation. We first convert Eq.(\ref{1})
into its matrix version:
\begin{eqnarray}
{\dot
S}_n=-i\frac{[S_n,S_{n+1}]}{2}-i\frac{[S_n,S_{n-1}]}{2},\label{matrix}
\end{eqnarray}
where $S_n=\left(\begin{array}{cc}
     s_n^1 & s_n^2-is_n^3\\
     s_n^2+is_n^3 & -s_n^1
     \end{array}
     \right)$. Then we define a discrete connection $\{A_n\}$ by
\begin{eqnarray}
A_n=(L_n,M_n),\label{conn}
\end{eqnarray}
where $L_n=\frac{z+z^{-1}}{2}I+\frac{z-z^{-1}}{2}S_n$,
$M_n=i\left(1-\frac{z^2+z^{-2}}{2}\right)\frac{S_n+S_{n-1}}{2}-i\frac{z^2-z^{-2}}{2}
     \frac{I+S_{n-1}S_n}{2}$ with $I$ denoting the $2\times 2$ identity matrix
and $z$ being a free spectral parameter. It is a direct and lengthy
computation, in the similar way displayed in the appendix of
\cite{dq6}, that the corresponding discrete curvature $\{F_n^A\}$ is
given by $F^A_{n}:=\dot
L_n-M_{n+1}L_n+L_nM_n=\frac{z-z^{-1}}{2}\left({\dot
S}_n+i\frac{[S_n,S_{n-1}]}{2}+i\frac{[S_n,S_{n+1}]}{2}\right)+i\frac{-z^3+z+z^{-1}-z^{-3}}{4}({\bf
S}_n\cdot{\bf S}_{n-1}-{\bf S}_n\cdot{\bf
S}_{n+1})S_n+i\frac{-z^3+z-z^{-1}+z^{-3}}{4}({\bf S}_n\cdot{\bf
S}_{n-1}-{\bf S}_n\cdot{\bf S}_{n+1})I$. Here we have used the
identity: $S_{n+1}S_n=-S_nS_{n+1}+({\bf S}_n\cdot{\bf S}_{n+1})2I$
and similarly $S_{n}S_{n-1}=-S_{n-1}S_{n}+({\bf S}_n\cdot{\bf
S}_{n-1})2I$ in the computation. Thus we see that, if we set
$K_n=i\frac{-z^3+z+z^{-1}-z^{-3}}{4}({\bf S}_n\cdot{\bf
S}_{n-1}-{\bf S}_n\cdot{\bf
S}_{n+1})S_n+i\frac{-z^3+z-z^{-1}+z^{-3}}{4}({\bf S}_n\cdot{\bf
S}_{n-1}-{\bf S}_n\cdot{\bf S}_{n+1})I$, Eq.(\ref{matrix}) is
equivalent to holding the following prescribed discrete curvature
representation:
\begin{eqnarray}
F^A_n:=\dot L_n-M_{n+1}L_n+L_nM_n=K_n,\quad \forall n. \label{2}
\end{eqnarray}

Our aim in this section is to geometrically transform
Eq.(\ref{matrix}) (resp., the connection (\ref{conn})) to a discrete
nonlinear Schr\"odinger-like equation (resp., a new connection) by a
discrete gauge sequence $\{G_n\}$. So the key point is to find such
a $\{G_n\}$. For this purpose, for a matrix sequence $\{S_n\}$
solving Eq.(\ref{matrix}), we shall find $\{G_n\}$ to satisfy
\begin{eqnarray}
\sigma_3=-G_nS_nG_n^{-1},\quad G_{n+1}=\left(\begin{array}{cc}
     1& \bar q_n\\
     -q_n & 1
     \end{array}
     \right)G_n ~~\hbox{for~some~(complex~valued)}~~q_n,\label{3}
\end{eqnarray}
where $\sigma_3=\left(\begin{array}{cc}
     1 & 0\\
     0& -1
     \end{array}
     \right)$ is the Pauli matrix. The first equation of (\ref{3})
has a class of solutions of the form
\begin{eqnarray}
G_n=\frac{i}{\sqrt{2(1-s_n^1)}}(\sigma_3-S_n)\diag(F_n,\bar F_n),
\label{Gn}
\end{eqnarray}
where $\{F_n\}$ is free and to be specified latter. It is a
straightforward verification that, in order for (\ref{Gn}) to
fulfill the second equation of (\ref{3}) for some complex sequence
$\{q_n\}$, $F_n$ is forced to satisfy the following iterated
relation
$$
(1-s_n^1)(1-s_{n+1}^1)\overline{F_{n+1}F^{-1}_n}+(s_n^2-is_n^3)(s_{n+1}^2+is_{n+1}^3)
F_{n+1}F^{-1}_n=2\sqrt{(1-s_n^1)(1-s_{n+1}^1)},\quad \forall n.$$ So
we may determine $F_n$ progressively by this relation in $n$ and
hence have proved the existence of a desired gauge sequence
$\{G_n\}$.

For the connection (\ref{conn}), we use $\{G_n\}$ to define a new
connection $\{A^G_n=(L_n^G(t,z),M_n^G(t,z))\}$ by
$L_n^G(t,z)=G_{n+1}L_nG^{-1}_n=G_{n+1}\left(\frac{z+z^{-1}}{2}I+\frac{z-z^{-1}}{2}S_n\right)G_n^{-1}
=\left(\begin{array}{cc}
     z^{-1} & z\bar q_n\\
     -z^{-1} q_n& z
     \end{array}
     \right)$ and $M_n^G(t,z)$ $={\dot G}_nG_n^{-1}+G_nM_nG^{-1}_n={\dot G}_nG_n^{-1}+
     i\left(\begin{array}{cc}
     (-1+z^{-2})\frac{1}{1+|q_{n-1}|^2} & (1-z^2)\frac{\bar q_{n-1}}{1+|q_{n-1}|^2}\\
     (1-z^{-2})\frac{q_{n-1}}{1+|q_{n-1}|^2}& (1-z^2)\frac{1}{1+|q_{n-1}|^2}
     \end{array}
     \right)$. Here we have used the relations (\ref{3}) in the
computation. Since $L_n,M_n$ satisfy the prescribed discrete
curvature representation (\ref{2}), from Lemma 1 proved in
\cite{dq6} we know that $L^G_n,M^G_n$ should fulfill
\begin{eqnarray}
\dot L_n^G-M_{n+1}^GL_n^G+L_n^GM_n^G=G_{n+1}K_nG^{-1}_n, ~~\forall
n. \label{5}
\end{eqnarray}
It is a direct computation that
\begin{eqnarray}
 &&\hbox{l.h.s. of ~(\ref{5})}=\left(\begin{array}{cc}
   1 & \bar {q}_{n}\\
   -{q}_{n} &1
                                 \end{array}
                           \right)\left[\left(\begin{array}{cc}
   z^{-1} & 0\\
   0 &z
                                 \end{array}
                           \right)\dot{ G}_n{
                           G}^{-1}_n-\dot{ G}_n{G}^{-1}_n
\left(\begin{array}{cc}
   z^{-1} & 0\\
   0 &z
                                 \end{array}
                           \right)\right]\nn\\
&&+i\left(\begin{array}{cc}
   a_n& -\frac{(z^{-1}-z^{3})\bar q_n}{1+|q_n|^2}+\frac{(z-z^3)\bar q_n+(z^{-1}-z)\bar q_{n-1}}
   {1+|q_{n-1}|^2}\\
   -\frac{(z-z^{-3})q_n}{1+|q_n|^2}+\frac{(z^{-1}-z^{-3})q_n+(z-z^{-1}) q_{n-1}}
   {1+|q_{n-1}|^2}& b_n
                                 \end{array}
                           \right)\label{1h}
                           \end{eqnarray}
where
$a_n=\frac{(z^{-1}-z^{-3})-(z-z^{-1})|q_n|^2}{1+|q_n|^2}+\frac{(-z^{-1}+z^{-3})+(z-z^{-1})\bar
q_nq_{n-1}}
   {1+|q_{n-1}|^2}$ and $b_n=-\frac{(z-z^3)+(z-z^{-1})|q_n|^2}{1+|q_n|^2}+\frac{(z-z^{3})+(z-z^{-1})q_n\bar q_{n-1}}
   {1+|q_{n-1}|^2}$ and
\begin{eqnarray}\hbox{r.h.s. of
~(\ref{5})}=i\left(\frac{1}{1+|q_n|^2}-\frac{1}{1+|q_{n-1}|^2}\right)\left(\begin{array}{cc}
   z^{-1}-z^{-3} & (-z+z^3)\bar q_n\\
   (-z^{-1}+z^{-3})q_n&-z+z^3
                                 \end{array}
                           \right).\label{2h}
\end{eqnarray}
Here we have used the relation (\ref{3}) and some identities
displayed in the appendix of \cite{dq6}), e.g., ${\bf S}_n\cdot{\bf
S}_{n+1}=\frac{1-|q_n|^2}{1+|q_n|^2}$ and ${\bf S}_n\cdot{\bf
S}_{n-1}=\frac{1-|q_{n-1}|^2}{1+|q_{n-1}|^2}$, in the above
computations. Thus, by substituting (\ref{1h}) and (\ref{2h}) into
(\ref{5}), the equation of the off-diagonal part of ({\ref{5}) leads
to
\begin{eqnarray}
\dot{ G}_n{ G}^{-1}_n=\left(\begin{array}{cc}
    *& i\frac{\bar q_n}{1+|q_n|^2}-i\frac{\bar q_{n-1}}{1+|q_{n-1}|^2}\\
   i\frac{ q_n}{1+|q_n|^2}-i\frac{ q_{n-1}}{1+|q_{n-1}|^2} &*
                                 \end{array}
                           \right),\label{85}
\end{eqnarray}
where $*$ are some expressions which cannot be carried out at this
moment. On the other hand, at the same time we also have
\begin{eqnarray}
&&\hbox{l.h.s. of (\ref{5})}=\left(\begin{array}{cc}
   0 & \dot {\bar q}_{n}
   z\\
   -{\dot q}_{n}z^{-1} &0
                                 \end{array}
                           \right)+\left(\begin{array}{cc}
   z^{-1} & \bar { q}_{n}z\\
   -{ q}_{n}z^{-1} &z
                                 \end{array}
                           \right)\dot{ G}_n{
                           G}^{-1}_n\nn\\&&~~~~~~~~~~~~~~~~~~~-\dot{ G}_{n+1}{
                           G}^{-1}_{n+1}
\left(\begin{array}{cc}
   z^{-1} & \bar {q}_{n}z\\
   -{ q}_{n}z^{-1}&z
                                 \end{array}
                           \right)\nn\\
&&+i\left(\begin{array}{cc}
   a_n& -\frac{(z^{-1}-z^{3})\bar q_n}{1+|q_n|^2}+\frac{(z-z^3)\bar q_n+(z^{-1}-z)\bar q_{n-1}}
   {1+|q_{n-1}|^2}\\
   -\frac{(z-z^{-3})q_n}{1+|q_n|^2}+\frac{(z^{-1}-z^{-3})q_n+(z-z^{-1}) q_{n-1}}
   {1+|q_{n-1}|^2}& b_n
                                 \end{array}
                           \right).\label{3h}
\end{eqnarray}
By substituting (\ref{3h}) and (\ref{2h}) into the equation
(\ref{5}) and combining with (\ref{85}), the diagonal part in this
time implies
\begin{eqnarray}
\dot{G}_n{G}^{-1}_n=\left(\begin{array}{cc}
    \al_n& i\frac{\bar q_n}{1+|q_n|^2}-i\frac{\bar q_{n-1}}{1+|q_{n-1}|^2}\\
   i\frac{ q_n}{1+|q_n|^2}-i\frac{ q_{n-1}}{1+|q_{n-1}|^2}
     &\sigma_n
                                 \end{array}
                           \right)\label{6}
\end{eqnarray}
where $\al_n$ and $\sigma_n$ satisfy
\begin{eqnarray}
\left\{\begin{array}{c}\al_n-\al_{n+1}=i\left(\frac{\bar q_n
q_{n-1}}{1+|q_{n-1}|^2}-\frac{\bar q_{n+1}q_n}{1+|q_{n+1}|^2}\right)\\
\sigma_n-\sigma_{n+1}=i\left(\frac{q_{n+1}\bar
q_n}{1+|q_{n+1}|^2}-\frac{q_n\bar
q_{n-1}}{1+|q_{n-1}|^2}\right).\end{array}\right.\label{alsi}
\end{eqnarray}
Obviously, the second equation of (\ref{alsi}) has a solution
$\sigma_n=\bar \al_n$. By substituting (\ref{6}) and (\ref{2h}) into
(\ref{5}), the off-diagonal part leads to $\{q_n\}$ satisfying the
following discrete nonlinear Schr\"odinger-like equation
\begin{eqnarray}
\dot
q_n+i\frac{q_{n+1}}{1+|q_{n+1}|^2}-2i\frac{q_n}{1+|q_n|^2}+i\frac{q_{n-1}}{1+|q_{n-1}|^2}
+(\al_n-{\bar\al}_{n+1})q_n=0,\label{nonsch}
\end{eqnarray}
where $\{\al_n\}$ solves the first equation of (\ref{alsi}). This
shows that the DHF (\ref{1}) is the (discrete) gauge equivalent to
the discrete nonlinear Schr\"odinger-like equation (\ref{nonsch}).

We would point out that Eq.(\ref{nonsch}) sets
$\{A^G_n=(L_n^G(t,z),M_n^G(t,z))\}$ as its discrete connection and
has prescribed discrete curvature representation: $F^G_n=K^G_n$,
where $K^G_n$ is given by right-hand-side of (\ref{2h}). Notice that
the expression $K^G_n$ is to be zero at $z=1$, it can also be proved
conversely that the discrete nonlinear Schr\"odinger-like equation
(\ref{nonsch}) is (discrete) gauge equivalent to the DHF (\ref{1})
by the gauge sequence $\{G_n\}$ satisfying $G_{n+1}=L^G_n(t,1)G_n$,
$\dot G_n=M^G(t,1)G_n$, $\forall n$ (the existence of such a
sequence $\{G_n\}$ is because of the integrability of $\{A^G_n\}$,
or in other words its zero curvature representation, at $z=1$). The
details are omitted here and one may refer to \cite{dq6} for a
reference.

$\{\al_n\}$ appeared in Eq.(\ref{nonsch}) can be solved out from the
first equation of (\ref{alsi}) as follows in different two
approaches. One is obtained by iterating from initial data $q_0$,
$q_{-1}$ and $\al_0$:
\begin{eqnarray}
\al_n=\left\{\begin{array}{c} i\frac{\bar q_n
q_{n-1}}{1+|q_{n}|^2}+i\sum_{k=1}^n\bigg(\frac{1}{1+|q_{k-1}|^2}-\frac{1}{1+|q_{k-2}|^2}\bigg)\bar
q_{k-1}q_{k-2}-i\frac{\bar q_0q_{-1}}{1+|q_{-1}|^2}+\al_0,~~n>0\\
\al_0,\qquad\qquad \qquad\qquad\qquad\qquad\qquad\qquad \qquad\qquad \qquad\qquad \qquad\quad ~n=0\\
i\frac{\bar q_n
q_{n-1}}{1+|q_{n}|^2}+i\sum_{k=n}^{-1}\bigg(\frac{1}{1+|q_{k}|^2}-\frac{1}{1+|q_{k+1}|^2}\bigg)\bar
q_{k+1}q_{k}-i\frac{\bar q_0q_{-1}}{1+|q_{0}|^2}+\al_0,~~~~~~~n<0.
\end{array}\right.\label{al1}
\end{eqnarray}
The other is obtained by iterating from the boundary data at
positive infinity:
\begin{eqnarray}
\al_{n}=
i\sum_{k=n}^{+\infty}\left({{1}\over{1+|q_k|^2}}-{{1}\over{1+|q_{k+1}|^2}}
\right)\bar q_{k+1} q_k+i{{\bar q_n
q_{n-1}}\over{1+|q_{n-1}|^2}}\quad \forall n. \label{al2}
\end{eqnarray}
This reflects that the property of the nearest neighbor exchange
interaction for Eq.(\ref{1}) collapses for its gauged equivalent
equation (\ref{nonsch}), since the $n$-th chain's exchange
interaction in Eq.(\ref{nonsch}) relates to chains from $-1$-st
chain to $(n\pm1)$-th chain or all the chains with label $\ge n-1$.
Though this implies that the dynamics of Eq.(\ref{nonsch}) and hence
the original Eq.(\ref{1}) will be complicated, the exposition of the
approximate linear stability in the next section for the stationary
version of Eq.(\ref{nonsch}) shows in interesting fact: the $n$-th
chain term is related to its nearest neighbor exchange interaction
in the stability analysis. This reflects that Eq.({\ref{nonsch})
does have the property of the nearest neighbor exchange interaction
in some sense. We note that (\cite{Ish,dq4,dq6}) the discrete
nonlinear Schrodinger equations without nonconstant denominator
terms (e.g. the AL equation: $i\dot
q_n+(q_{n+1}+q_{n-1}-2q_n)+|q_n|^2(q_{n+1}+q_{n-1})=0$) are gauge
equivalent to the (modified) discrete Heisenberg spin chain models
with nonconstant denominator terms (e.g., the Ishimori equation: $
\dot{ S}_n=2\frac{{ S}_{n+1}\times{ S}_n}{1+{\bf S}_{n+1}\cdot{\bf
S}_{n}}+2\frac{{S}_{n}\times{ S}_{n-1}]}{1+{\bf S}_n{\cdot}{\bf
S}_{n-1}}$). And, meanwhile, the above exposition indicates that the
discrete Heisenberg spin chain model (\ref{1}) without nonconstant
denominator terms is gauge equivalent to the discrete nonlinear
Schrodinger-type equation (\ref{nonsch}) with nonconstant
denominator terms. This is an interesting duality phenomenon between
discrete nonlinear Schrodinger-type equations and discrete
Heisenberg spin chain models.

Before we use (\ref{nonsch}) to study the transmission problem of
the DHF, let's give a general description of constructing solutions
to the DHF (\ref{1}), or equivalently (\ref{matrix}), from those to
the discrete nonlinear Schr\"odinger-like equation (\ref{nonsch}).
For a solution $\{q_n\}$ to Eq.(\ref{nonsch}), let $G_0$ be a
fundamental solution to
\begin{eqnarray}
\dot G_0=M_0(t,1)G_0=i\left(\begin{array}{cc}
    \al_0& \frac{\bar q_0}{1+|q_0|^2}-\frac{\bar q_{-1}}{1+|q_{-1}|^2}\\
   \frac{ q_0}{1+|q_0|^2}-\frac{ q_{-1}}{1+|q_{-1}|^2}
     &\bar \al_0
                                 \end{array}
                           \right)G_0. \label{G0}
\end{eqnarray}
It is a direct verification that, by successive iteration,
\begin{eqnarray}
G_n&=&\left\{\begin{array}{c}L^G_{n-1}(t,1)G_{n-1}, ~~~~n>0\\
L^G_{n+1}(t,1))^{-1}G_{n+1},~
n<0\end{array}\right.=\left\{\begin{array}{c}L^G_{n-1}(t,1)\cdots
L^G_0(t,1)G_0,\qquad ~~n>0\\
(L^G_{n+1}(t,1)\cdots L^G_0(t,1))^{-1}G_0,\quad
n<0\end{array}\right.
 \label{40}
\end{eqnarray}
solves $G_{n+1}=L^G_n(t,1)G_n$, $\dot G_n=M^G(t,1)G_n$, $\forall n$.
Therefore $ \{S_n=-G^{-1}_n\sigma_3G_n\}$ is a solution to the DHF
(\ref{matrix}) which corresponds to $\{q_n\}$ under the discrete
gauge transformation.

\section*{\S3 Applications} In this section, we shall use the gauged
equivalent equation (\ref{nonsch}) to study dynamical properties of
the discrete Heisenberg ferromagnetic spin chain (\ref{1}). We
mainly focus on exploring the transmission property of the DHF
(\ref{1}) and its related linear stability analysis for the
stationary version of the equation (\ref{nonsch}).

\subsection*{{\bf A}  Transmission properties} In this subsection, we study
as a physical application whether the wave transmission property
(\cite{DLS,HSGT}) of the discrete nonlinear Schr\"odinger-like
equation (\ref{nonsch}) and transfer it to that of  the DHF
(\ref{1}) under the action of discrete gauge transformations. In
order to do these,  let's consider the following recurrence equation
originated from the discrete nonlinear Schr\"odinger-like equation
(\ref{nonsch}) by setting $q_n(t)={\varphi}_n\exp(-iEt)$:
\begin{eqnarray}
-E {\varphi}_n
+\frac{{\varphi}_{n+1}}{1+|{\varphi}_{n+1}|^2}-2\frac{{\varphi}_n}{1+|{\varphi}_n|^2}
+\frac{{\varphi}_{n-1}}{1+|{\varphi}_{n-1}|^2}
-i(\al_n-{\bar\al}_{n+1}){\varphi}_n=0,\label{mnonsch}
\end{eqnarray}
where $\varphi_n$ is a complex amplitude independent of the time
variable $t$, $E$ is a real parameter and $\al_n$ independent of the
time variable $t$ solves the first equation of (\ref{alsi}), which
reads from (\ref{al2}),
\begin{eqnarray}
\al_{n}=
i\sum_{k=n}^{+\infty}\left({{1}\over{1+|\varphi_k|^2}}-{{1}\over{1+|\varphi_{k+1}|^2}}
\right)\bar\varphi_{k+1} \varphi_k+i{{\bar\varphi_n
\varphi_{n-1}}\over{1+|\varphi_{n-1}|^2}}\quad \forall n.
\label{30}
\end{eqnarray}

We thus study the transmission problem of Eq.(\ref{mnonsch}) via a
similar numerical way of the AL-DNLS lattice chain displayed in
\cite{DLS,HSGT} (please refer to their papers for details).
First we note that Eq.(\ref{mnonsch}) has a stationary solution as
follows:
\begin{eqnarray}
\varphi_n=Te^{i\kappa n},\quad n\in {\bf Z} \label{stable}
\end{eqnarray}
where the real parameter $E$  satisfies the consistent condition
\begin{eqnarray}2\cos
\kappa={{2+E(1+T^2)}\over{1+T^2}}.\label{kappa}
\end{eqnarray}
We now consider the problem: A finite nonlinear segment $0\le n\le
N-1$ of length $N$ in the nonnegative stationary regime is embedded
in a nonlinear chain $\{\varphi_n\}_{n\in{\bf Z}}$ satisfying
(\ref{mnonsch}) with
\begin{eqnarray}
\varphi_n=\left\{\begin{array}{c}
R_0e^{i\kappa}+ Re^{-i\kappa},~\quad
n=0\\
Te^{i\kappa n},~~~\qquad\qquad
n\ge N,
\end{array}\right.\label{80}
\end{eqnarray}
where (\ref{kappa}) is fulfilled. This can be regarded as that an
incident plane wave $R_0e^{i\kappa}$ on the left ($n=0$) induces a
reflected plane wave $Re^{-i\kappa}$ on the left and a transmitted
plane wave $Te^{i\kappa n}$ on the right ($n\ge N$). $R_0$ is called
the amplitude of the incoming wave, $R$ the amplitude of the
reflected waves and $T$ the transmitted amplitude at the right end
of the nonlinear chain; $\kappa$ is called the out-coming wave
number;  $|R_0|^2$ and $|T|^2$ are also called the in-coming and the
transmitted intensity respectively. The medium is completely
nonlinear, thus the transmission coefficient $T$ as a function of
$R_0$ may not uniquely determined. If this is true, according to the
sense made in \cite{Fl,DLS}, it is called bistability. For the
forward transmission problem of the AL-DNLS lattice chain, there
occurs exactly the bistability phenomenon (\cite{DLS,HSGT}), which
leads Delyon et al \cite{DLS} to consider the backward transmitted
problem for the DNLS. They proved that the pair $(\kappa, |T|)$
initializes the incident intensity $|R_0|$ completely and then
displayed the transmission behavior of the DNLS in the
$(\kappa,|T|)$ plane.

Now we consider the similar backward transmission problem for
solutions with the type (\ref{80}) to the present equation
(\ref{mnonsch}). Following Wan and Soukoulis \cite{WS}, we use polar
coordinates for $\varphi_n$, that is $\varphi_n=R_n e^{i\theta_n}$.
For a given pair $(\kappa,|T|)$ (without loss of the generality, we
may assume $T>0$ in this paper), we try to iterate
Eq.(\ref{mnonsch}) from $n=N$ to $-\infty$ successively and to
determine the amplitudes $(R_{N-1},\cdots,R_0)$ and phases
$(\theta_{N-1},\cdots,\theta_0)$. The existence of solutions with
the type (\ref{80}) to Eq.(\ref{mnonsch}) will be shown below in the
algorithms. In order to further support the existence of such
solutions, an approximate linear stability analysis around the the
stationary solution (\ref{stable}) will be given in the next
section, which further reveals the linear stability behaviors of the
equation ({\ref{mnonsch}). In this iterating process, we find that
there are two choices in determining the amplitudes
$(R_{N-1},\cdots,R_0)$ in each step. This implies that, not like the
transmission problem of the AL-DNLS, the pair $(\kappa, T)$ does not
initialize the incident intensity $|R_0|$ completely, or in other
words, the backward transmission problem for the equation
(\ref{mnonsch}) with solutions of type (\ref{80}) is still bistable.

We first introduce a {\bf definite algorithm} such that the pair
($\kappa$,$T$)  initializes the amplitudes $(R_{N-1},\cdots,R_0)$
and phases $(\theta_{N-1},\cdots,\theta_1)$ completely. In fact,
such an {algorithm} is designed as follows.
 From
$\varphi_{n}=Te^{i\kappa n},~n\ge N$ we get $\al_n=i{{\bar\varphi_n
\varphi_{n-1}}\over{1+|\varphi_{n-1}|^2}},~n\ge N$ from (\ref{30})
and the equation (\ref{mnonsch}) for $n\ge N+1$ is now equivalent to
(\ref{kappa}). The equation (\ref{mnonsch}) for $n=N$ is rewritten
as
\begin{eqnarray}
\left\{\begin{array}{c}T\cos(\kappa)+\frac{1+T^2}{1+R_{N-1}^2}R_{N-1}\cos(\Delta\theta_{N})
={{2+E(1+T^2)}\over{1+T^2}}T,\\
T\sin(\kappa)-\frac{1+T^2}{1+R_{N-1}^2}R_{N-1}\sin(\Delta\theta_N)=0,~~~~~~~~~~~
~~\end{array}\right.\nn
\end{eqnarray}
where $\Delta \theta_n=\theta_n-\theta_{n-1}$ (we remark that
$\theta_N=\kappa N$), which leads $R_{N-1}$ to satisfying
$T=\frac{1+T^2}{1+R_{N-1}^2}R_{N-1}$ and $\theta_{N-1}=\kappa(N-1)$.
Since the quadratic equation $T=\frac{1+T^2}{1+x^2}x$ in $x$ just
has two self-reciprocal solutions: $x=T$ and $x=\frac{1}{T}$, we
then determine $R_{N-1}$ according to {\bf the rule}: to choose
$R_{N-1}>1$ when $0<T<1$ and $R_{N-1}<1$ when $T>1$ in order to
avoid getting the stationary solution (\ref{stable}). Next, from
(\ref{30}) we see that, for $n<N$,
$\al_n=i\sum_{k=n+1}^N\bigg(\frac{1}{1+|\varphi_{k-1}|^2}-\frac{1}{1+|\varphi_k|^2}
\bigg)\bar\varphi_k\varphi_{k-1}+i\frac{\bar\varphi_n\varphi_{n-1}}{1+|\varphi_{n-1}|^2}$.
Then the equation (\ref{mnonsch}) for $n<N$ is is equivalent to
\begin{eqnarray}
\left\{\begin{array}{c}\frac{R_{n-1}}{1+R^2_{n-1}}\cos(\Delta\theta_n)=
\frac{R_{n}}{1+R^2_{n}}\Bigg\{\frac{E(1+R^2_n)+2}{1+R_{n}^2}
-\frac{R_{n+1}(\frac{1}{R_n}-R^3_n+2R_n+2R_nR^2_{n+1})}{(1+R^2_n)(1+R^2_{n+1})}\cos(\Delta\theta_{n+1})\\
\qquad\qquad\qquad-2\sum_{k=n+1}^{N-1}\left(\frac{1}{1+R_{k}^2}-\frac{1}{1+R_{k+1}^2}
\right)R_kR_{k+1}\cos(\Delta\theta_{k+1})\Bigg\},\\
\frac{R_{n-1}}{1+R^2_{n-1}}\sin(\Delta\theta_n)=\frac{R_{n+1}}{1+R^2_{n+1}}\cos(\Delta\theta_{n+1}).~~~~~~~~~~~
~~\qquad\qquad\qquad\qquad\qquad\qquad\end{array}\right.\label{31}
\end{eqnarray}
The above exposition indicates that $R_{n-1}$ satisfies
$$\frac{R_{n-1}}{1+R^2_{n-1}}=\sqrt{(\hbox{r.h.d~of~the~first
~eq.(\ref{31})})^2+(\hbox{r.h.d~of~the~second~eq.(\ref{31})})^2}$$
and thus also has two self-reciprocal solutions. Our {\bf rule} is
that in the second step and the following steps, we always choose
the root with $R_n>1$ for the quadratic equations. According to this
{algorithm}, we may uniquely determine the amplitudes and phases
from $n=N-1$ to $-\infty$ successively and especially the incident
intensity $|R_0|^2$. This procedure also indicates the existence of
solutions of type (\ref{80}) to the stationary equation
(\ref{mnonsch}).

If the resulting incoming wave intensity $|R_0|^2$ is of the same
order of $\frac{1}{|T|^2}$ (the reciprocal transmitted intensity
$|T|^2$) independent of $N$ when $0<T<1$, or is of the same order of
$|T|^2$ when $T>1$, we say that the nonlinear chain with wave number
$k$ and outgoing intensity $|T|^2$ is to be transmitting. If $R_0$
appears to be a rapidly increasing function of $N$, we say that this
nonlinear chain is to be non-transmitting. Figs.1(a) and (b) below
display the transmission behaviors in the $(k,T)$ parameter plane
with the chain length $N=110$ and 150, respectively, representing
region of transmitting (white) and non-transmitting (black)
behaviors.  In particular, Fig.1 (a) and (b) numerically show the
whole-scope of the transmitting behaviors for $\kappa\in(1,3.5)$ and
$T\in(0.5,2)$. The non-transmitting regions form some strange
patterns. The enlargement of one of these patterns, $(\kappa,T)\in
(1.3,1.7)\times(1,1.4)$, is further shown by Fig.1(c).  It is seen
from the figure that the twisting boundaries between the
transmitting region and the non-transmitting region in the
parameters plane always display spiral, cantor-like, and fractal
structure, which is very analogous to the structure of the standard
chaotic attractor. Furthermore, from the Fig.1 (a) and (b), we see
that the transmitting behaviors only depend seriously on the
parameters $\kappa$ and $T$ but do not depend sensitively on the
chain length $N$.

{\it Figure 1 around here. Please find it at the end of the paper.}

Besides the definite {algorithm}, we come to consider a {\bf
stochastic algorithm}. In such a stochastic algorithm, the
determination of the bi-valued $R_{N-1}$ should be the same as that
in the definite algorithm in order to avoid getting a stationary
solution; however, the latter determinations of $R_{n}$ ($0\le n\le
N-2$) in each step are based on the generation of a random number
$\xi$, i.e., $R_i=R_{-}$ if $\xi\ge c$; $R_i=R_{+}$ otherwise.
Naturally, our first choice of the generation of the random number
$\xi$ is the normal distribution and the value $c$ is taken as zero,
the expectation of this distribution. It could be seen from Fig.2
(a) that non-transmitting regions (black) periodically appear with
respect to $\kappa$, here the chain length is still to be $N=150$.
Also, this kind of phenomena could be somewhat found in Fig.1 (a)
and (b). Secondly, a Poisson distribution with an expectation $5$ is
adopted to produce the random number $\xi$ and $c$ is taken as $5$.
In this case, a similar periodic non-transmitting behavior is
displayed in Fig2.(b). Actually, thousands of numerical simulations
verify a fact that no matter what kind of stochastic distribution
and chain length are taken, the non-transmitting behavior is so
prevalent around the values $\kappa_k\approx 1.57+\frac{k\pi}{2}$
($k=0,1,\cdots$).  This periodic phenomena could be regarded as an
invariant non-transmitting behavior for Eq.(\ref{mnonsch}).

{\it Figure 2 around here. Please find it at the end of the paper.}

The transmission problem associated with the discrete Heisenberg
ferromagnetic spin chain (\ref{1}) (or equivalently (\ref{matrix}))
is now proposed as follows. There is a semi-infinite nonlinear
matrix wave chain $\{S_n=-G^{-1}_n\sigma_3G_n\}$  embedded in a
nonlinear chain of the DHF (\ref{matrix}),  where $G_n$  looks like
(from (\ref{40})):
\begin{eqnarray}
G_n=\left(\begin{array}{cc}
   1 & \bar q_{n-1}\\
   -q_{n-1} &1
                                 \end{array}
                           \right)\cdots\left(\begin{array}{cc}
   1 & \bar q_0\\
   -q_0 &1
                                 \end{array}
                           \right)G_0,~~n\ge 1\label{81}
\end{eqnarray}
with $G_0$ satisfying (\ref{G0}) and $q_n=\varphi_ne^{-iEt}$, where
$\varphi_n$ is independent of $t$ and given by (\ref{80}). This
nonlinear matrix wave chain is from the left towards the right,
where they are scattered into reflected and transmitted parts.
Similarly, $R_0, R$ present the amplitudes of the incoming and
reflected waves and $T$ the transmitted amplitude at the right end
of the nonlinear chain under consideration. $|R_0|^2$ and $|T|^2$
are also called the incoming wave intensity and the transmitted
intensity of the nonlinear matrix chain respectively. As we have
proved in the previous section that the DHF (\ref{1}) (i.e. the DHF
(\ref{matrix})) is the discrete gauge equivalent to the discrete
nonlinear Schr\"odinger-like equation (\ref{nonsch})  and verse
visa, we see that the transmission properties of the equation
(\ref{mnonsch}) just displayed, i.e., the above {\bf definite
algorithm} and {\bf stochastic algorithms}, are transferred
completely to the transmission problem (\ref{81}) of the DHF
(\ref{matrix}) under the action of discrete gauge transformations.
Thus the nonlinear matrix wave chain (\ref{81}) of the DHF
(\ref{matrix}) has the same transmission behaviors as those of
nonlinear chain (\ref{80}) of Eq.(\ref{mnonsch}). Fig.1(a),(b) and
(c) are completely suitable in displaying the transmission behaviors
of the finite nonlinear matrix chain of the DHF (\ref{matrix}) in
the parameters $(k,T)$ plane with chain length $N=110$ and 150 under
the definite {algorithm} and so do Fig.2(a) and (b)
under stochastic algorithms. Furthermore, from the Fig.1 (a) and
(b), we see that the transmitting behaviors for solutions
$\{S_n=-G^{-1}_n\sigma_3G_n\}$ with $\{G_n\}$ satisfying (\ref{81})
to the DHF (\ref{1}) only depend seriously on the parameters
$\kappa$ and $T$ but do not depend sensitively on the chain length
$N$. This shows that the important phenomenon of bistability occurs
still in spin magnetic physics. The periodic phenomenon of the
non-transmitting regions displayed in Fig.2 provides a new and
interesting transmission property of the DHF (\ref{1}).

\subsection*{{\bf B} Stability analysis}
In order to support the exposition in subsection A and show the
existence of stable solutions to (\ref{mnonsch}) around the
stationary solution ({\ref{stable}), we shall give an approximate
linear stability analysis for Eq.(\ref{mnonsch}). In what follows,
we focus on the locally linear stability of the stationary solution
(\ref{stable}) in the reverse iterative procedure through the linear
variational method \cite{Robin,Grimshaw}. For this purpose, we
naturally import the formulas as follows:
$$
  R_{n}=T+\tau_n, ~~\theta_n=n\kappa+\delta_n,~~ n=N, N-1, \cdots,
  2,1,
$$
where quantities $\tau_n$ and $\delta_n$, respectively, are regarded
as small perturbations of the modulus and argument of the stationary
solution (\ref{stable}).  Substitute the above formulas into Eq.
(\ref{31}) and then expand those nonlinear terms in the vicinity of
$T$ and $n\kappa$. A tedious calculation thus gives the following
equations at $O(\tau_i)$ and $O(\delta_i)$ ($i=n-1,n,n+1$):
\begin{equation}\label{iteration-linear}
 \left\{\begin{array}{lll}
     \frac{(1-T^2)\cos\kappa}{(1+T^2)^2}\tau_{n-1}&=&
    \frac{1-T^2}{(1+T^2)^2}\left[E+\frac{2}{1+T^2}-\frac{\cos\kappa}{1+T^2}
    -\frac{T^2\cos\kappa}{1+T^2}\right]\tau_n  \\
    & &  +\frac{T}{1+T^2} \left[-\frac{4T}{(1+T^2)^2}+\frac{\cos\kappa}{T(1+T^2)}
        -\frac{2T(1-T^2)\cos\kappa }{(1+T^2)^2}\right]\tau_n\\
    & &  +\frac{T}{1+T^2}\left[-\frac{(1-T^2)\cos\kappa }{T(1+T^2)^2}-
         -\frac{2T\cos\kappa}{1+T^2}+\frac{2T(1-T^2)\cos\kappa }{(1+T^2)^2}\right]\tau_{n+1}\\
    & &  +\frac{2T^4\cos\kappa}{(1+T^2)^2} \sum_{k=n+1}^{N-1} (\tau_k-\tau_{k+1}),\\

    \frac{\sin\kappa}{1+T^2}(\delta_{n}-\delta_{n-1})&=&
    -\frac{\sin\kappa}{1+T^2}(\delta_{n+1}-\delta_{n}),
    \\
     \frac{(1-T^2)\sin\kappa}{(1+T^2)^2}\tau_{n-1}&=&
    \frac{(1-T^2)\sin\kappa}{(1+T^2)^2}\tau_{n+1},\\
    \frac{T\cos\kappa}{1+T^2}(\delta_{n}-\delta_{n-1})&=&
    \frac{T\cos\kappa}{1+T^2}(\delta_{n+1}-\delta_{n}),
  \end{array}\right.
\end{equation}
where $n=N,N-1,\cdots,2,1$.  Clearly, combining the second and the
fourth equations in (\ref{iteration-linear}) leads to
$$
  \delta_{n+1}-\delta_{n}=0,
$$
which shows an identity of the series $\{\delta_n\}$ at
$O(\delta_n)$.  Hence, the linear stability of the argument of the
stationary solution in the reverse iteration procedure belongs to
the stable case. However, the nonlinear stability of this argument
becomes a critical case, depending on the further calculation at
$O(\delta_i^2)$. Due to the tediousness of notations, we omit the
calculation here.  Analogously, combining the first and the third
equations in (\ref{iteration-linear}) yields:
$$
  \tau_{n-1}=\left[\frac{1+T^4}{1-T^4}-\frac{2T^2}{(1-T^4)\cos\kappa}\right]\tau_n-\frac{2T^4}{1-T^4}\tau_N.
$$
Hence, it follows that for $n=N,\cdots,1$,
\begin{eqnarray}
  \tau_{n-1}&=&\Bigg\{\left[\frac{1+T^4}{1-T^4}-\frac{2T^2}{(1-T^4)\cos\kappa}\right]^{N-n+1}\nn\\&&-
             \frac{2T^4}{1-T^4}\sum_{j=0}^{N-n}
             \left[\frac{1+T^4}{1-T^4}-\frac{2T^2}{(1-T^4)\cos\kappa}\right]^{j}\Bigg\}\tau_N
             \triangleq \lambda_{N,n}\tau_N. \nn
\end{eqnarray}
Consequently, the linear stability of the modulus of the stationary
solution is stable in the reverse iteration provided that
$\lambda_{N,n}$ is uniformly bounded for all $N$ and $n$. More
precisely, the reverse stability holds when the following
inequalities are satisfied:
\begin{equation}\label{condition-reverse}
  \left|\frac{1+T^4}{1-T^4}-\frac{2T^2}{(1-T^4)\cos\kappa}\right|<1.
\end{equation}
In addition, notice that $\tau_n=\lambda^{-1}_{n,0}\tau_0$.  This
implies that the linear stability of the normal iteration procedure
is stable when $\lambda_{n,0}\not=0$ for all $n$.  Thus, this linear
stability is valid only provided
$$
  \kappa\not=\displaystyle\frac{k\pi}{2}~(k=0,\pm 1,\pm 2,\cdots).
$$
Obviously, this necessary stability condition is also included in
the condition (\ref{condition-reverse}).  As shown in Fig.3,
parameters selected from the black regions violate the inequality
(\ref{condition-reverse}).  Those diamond-like unstable regions
periodically appears in a period $2\pi$ with respect to the
parameter $\kappa$.

{\it Figure 3 around here. Please find it at the end of the paper.}

The relation of the linear stability analysis in this subsection and
the transmission exposition in the subsection A is that the
non-transmitting regions in Figure 1 or Figure 2 are always
contained in the unstable regions in Figure 3 (b). This also
indicates that the transmission exposition for solutions with the
type (\ref{80}) to Eq.(\ref{mnonsch}) is a more accurate description
of the linear stability analysis for Eq.(\ref{mnonsch}), which
exhibits some chaotic dynamics of the Eq.(\ref{nonsch}) and hence
those of the DHF (\ref{1}).

\section*{\S4 Conclusion}
We present a mechanism for displaying the transmission property of
the discrete Heisenberg ferromagnetic spin chain via a geometric
approach. More precisely, we first transform the DHF (\ref{1}) into
the discrete nonlinear Schr\"odinger-like equation (\ref{nonsch})
with the aid of the (discrete) Yang-Mills theory. Then we use the
stationary version of the discrete nonlinear Schr\"odinger-like
equation (\ref{nonsch}) to study the transmission problem of the DHF
(\ref{1}). In this procedure, we show that the determination of
transmitting coefficients in the backward transmission problem is
always bistable. Thus a definite algorithm and general stochastic
algorithms are presented. The corresponding transmitting behaviors
in the parameters $(\kappa,T)$ plane are shown by Figure 1 and 2. A
new invariant periodic phenomenon of the non-transmitting behavior
for the DHF, with a large probability, is revealed by an adoption of
various stochastic algorithms. By the way, an approximate linear
stability analysis for the stationary version of Eq.(\ref{mnonsch})
is also given for supporting the discussion of the transmission
expositions. We remark that though the gauged equivalent equation
(\ref{nonsch}) with (\ref{alsi}) looks much complicated and we
cannot give a direct physical application at present stage from it,
it is an effective way to utilize it to study some dynamical
properties of the DHF (\ref{1}). We believe that the discrete
nonlinear Schr\"odinger-like equation (\ref{nonsch}) can help us to
reveal much more and deeper quantum chaotic properties of the DHF
(\ref{1}).

\section* {Acknowledgments} This work was supported by the National
Natural Science Foundation of China (Grant Nos. 10531090, 10501008)
and STCSM.

\begin{figure}[htp]
\centering

\includegraphics[scale=0.5]{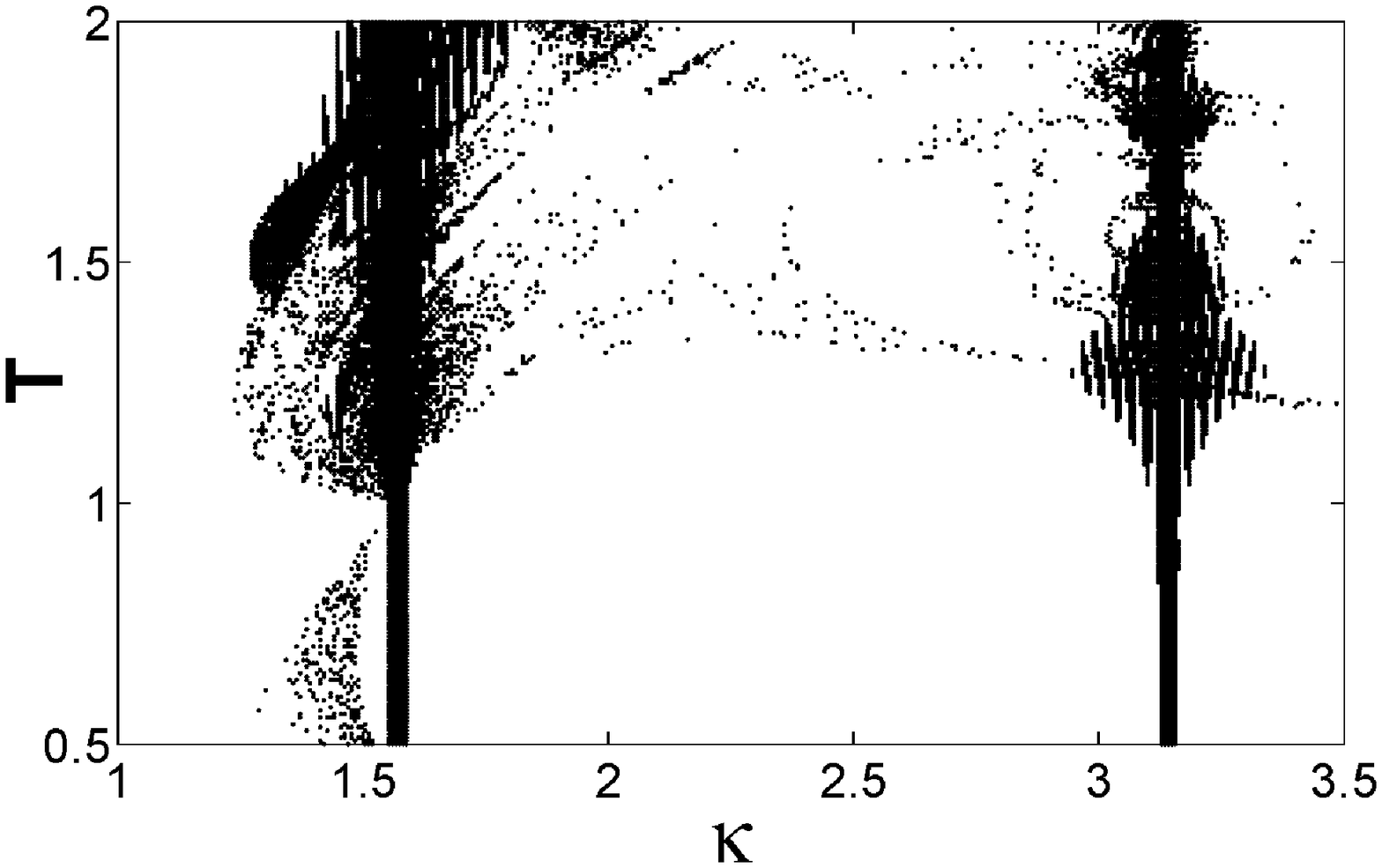}{\small(a)}
\includegraphics[scale=0.5]{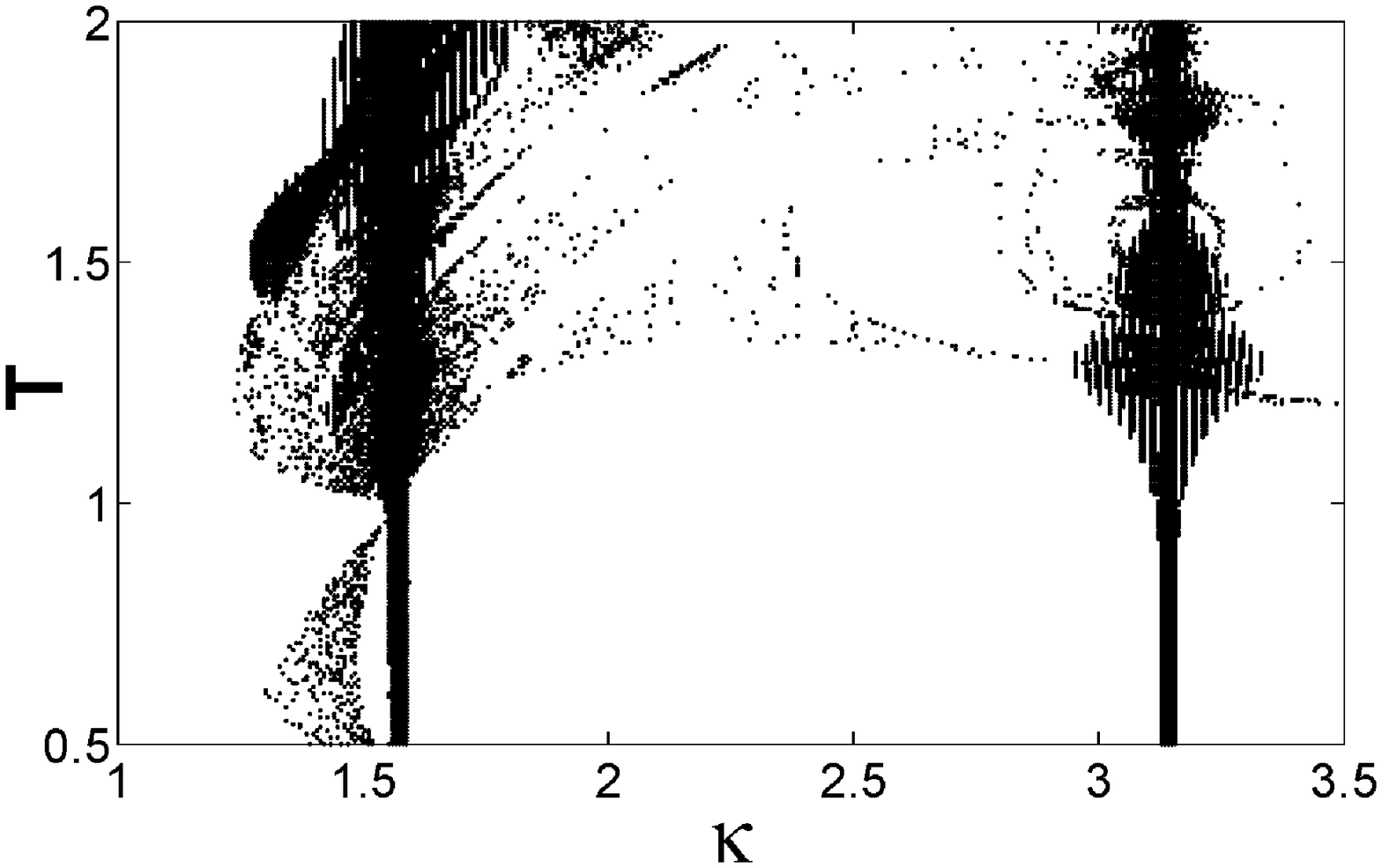}{\small(b)}
\includegraphics[scale=0.5]{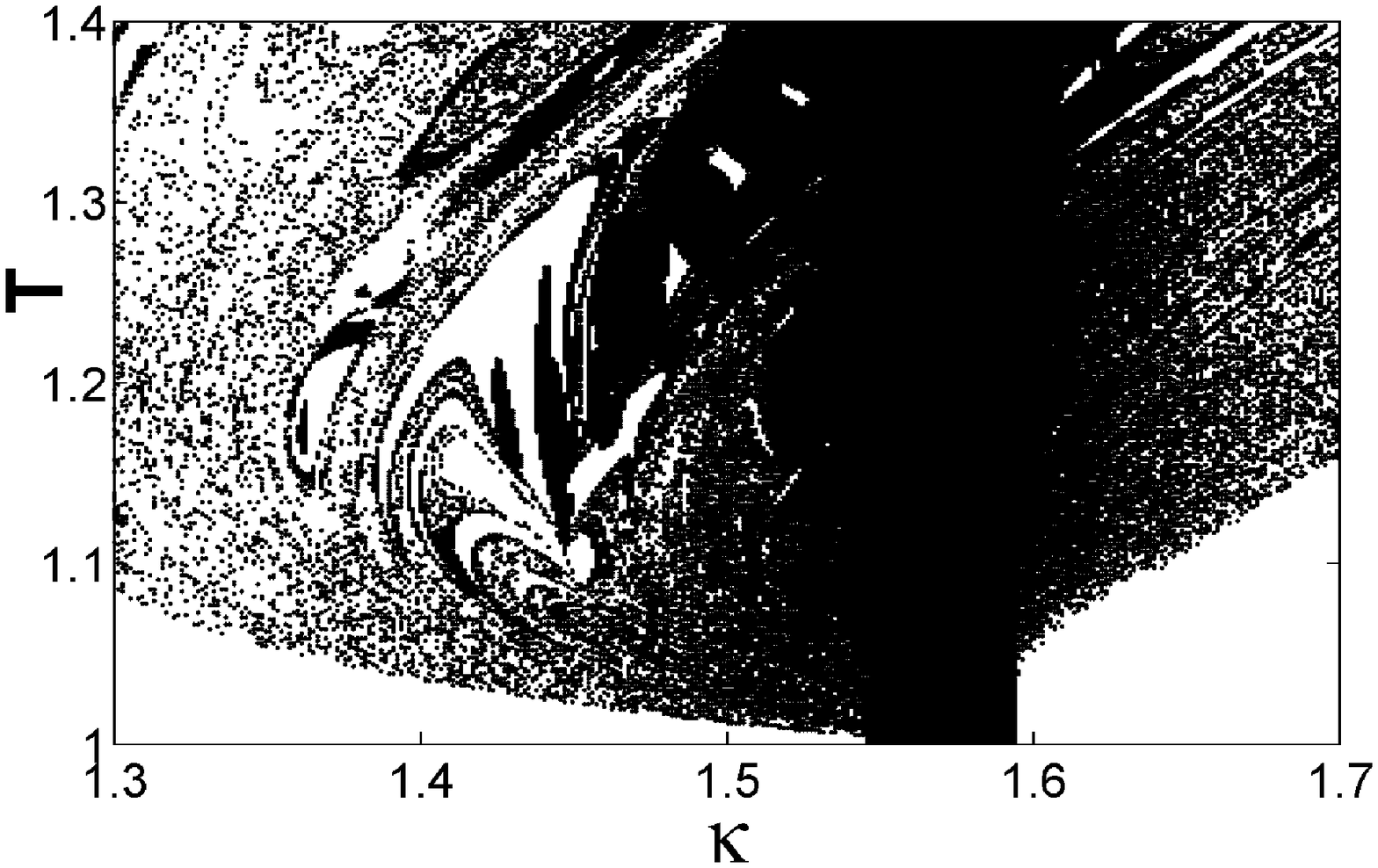}{\small(c)}

\renewcommand{\figurename}{Fig.}
\parbox[t]{12cm}{\caption{\small  The transmitting behavior in the $(\kappa,T)$
parameters plane $(\kappa,T)\in(1,3.5)\times(0.5,2)$ (a) with
$N=110$ and (b) with $N=150$, and in an enlarged region
$(\kappa,T)\in(1.3,1.7)\times(1,1.4)$ (c) with $N=150$. The
boundaries of transmitting (white) and non-transmitting (black)
regions exhibit fractal-like structure. .}}\label{fig1}
\end{figure}

\newpage

\begin{figure}[htp]

{\centering

\includegraphics[scale=0.5]{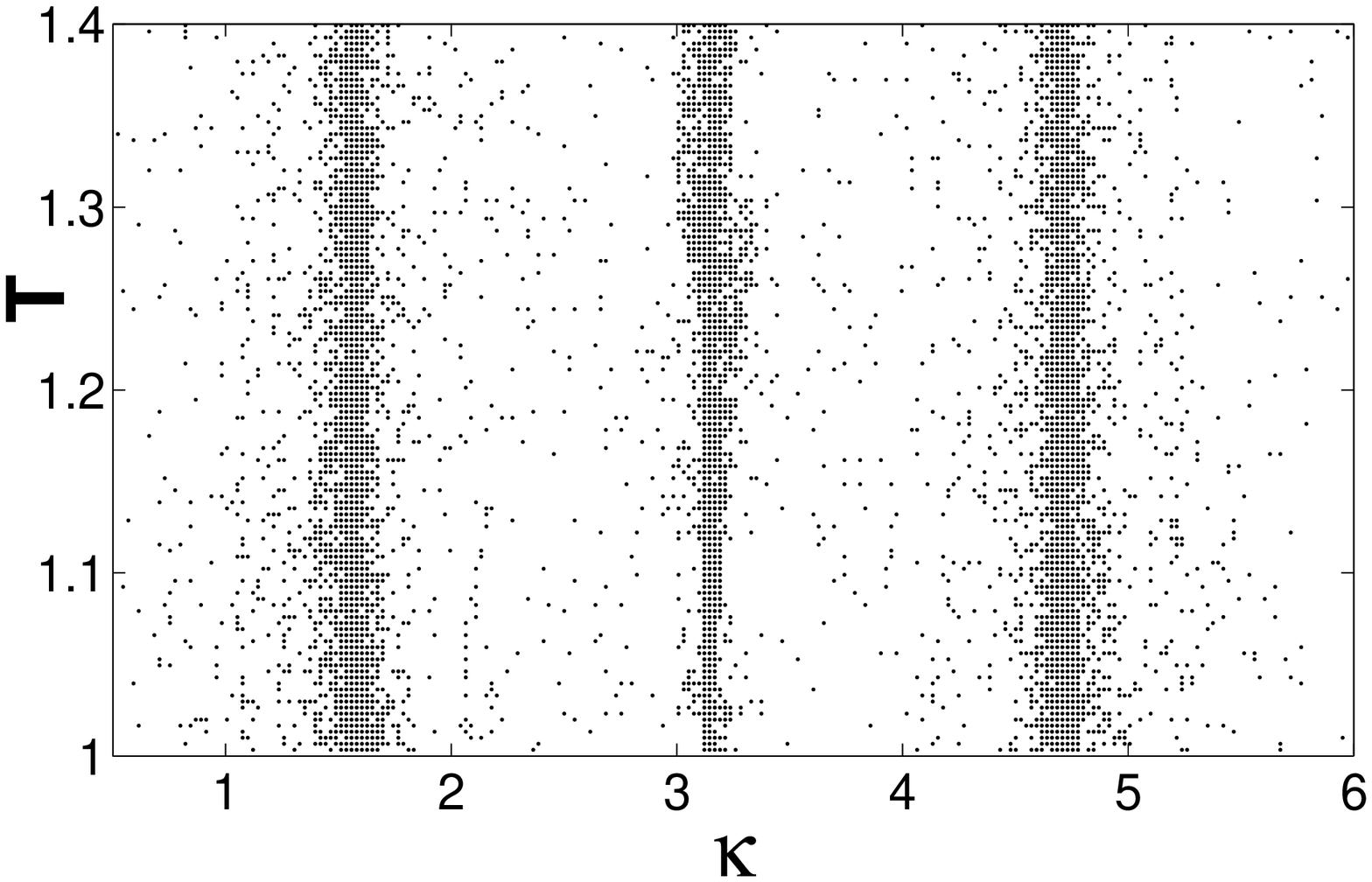}{\small(a)}
\includegraphics[scale=0.5]{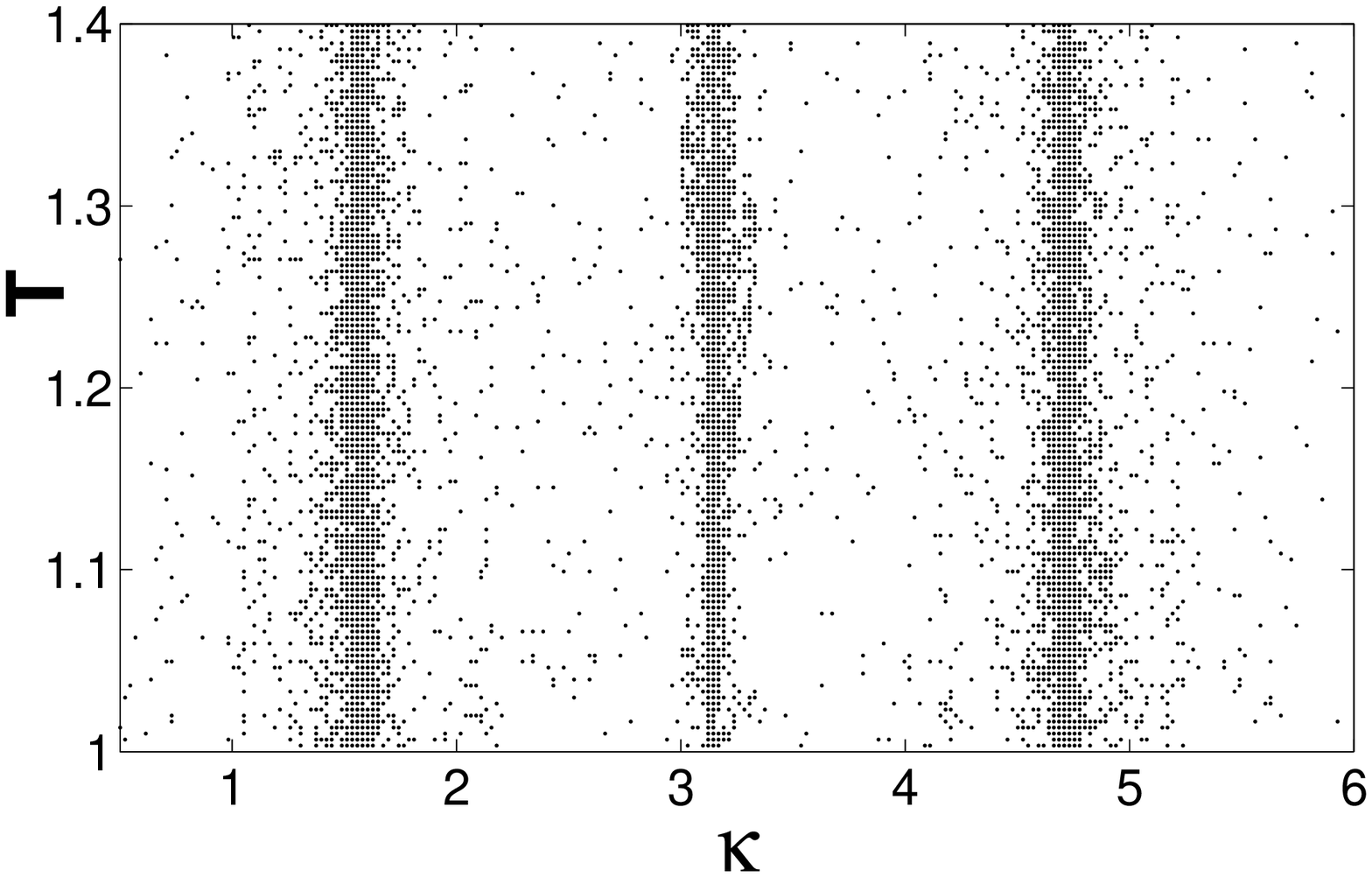}{\small(b)}

\renewcommand{\figurename}{Fig.}
\parbox[t]{12.6cm}{{\small  Fig.2.  The periodic phenomenon of the non-transmitting
behavior in $(\kappa,T)$ plane by adopting of the normal
distribution (a) and a Poisson distribution (b). Here the chain
length $N=150$. }}\label{fig2}

}
\end{figure}

\newpage

\begin{figure}[htp]

\begin{center}
\includegraphics[scale=0.5]{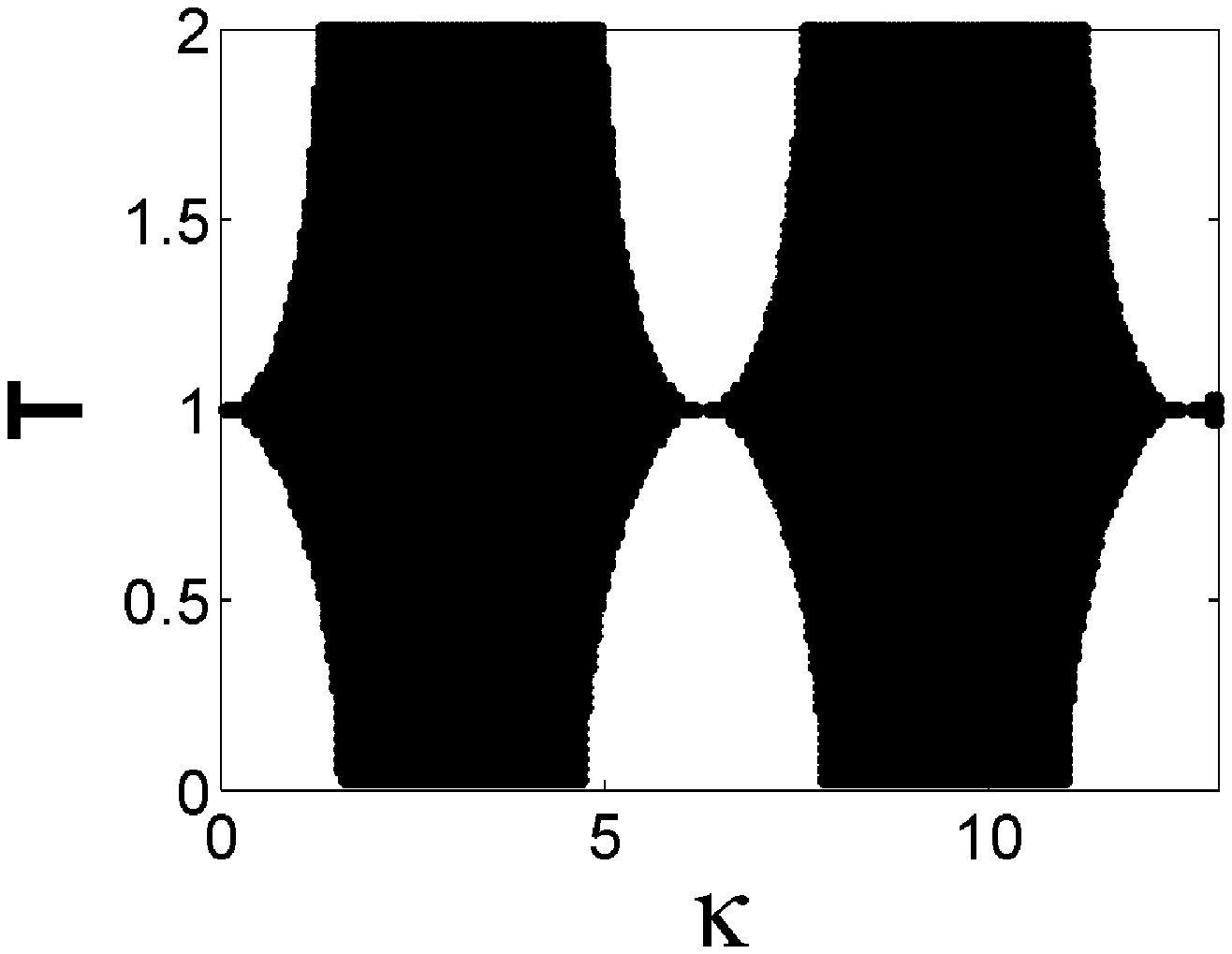}(a)
\includegraphics[scale=0.5]{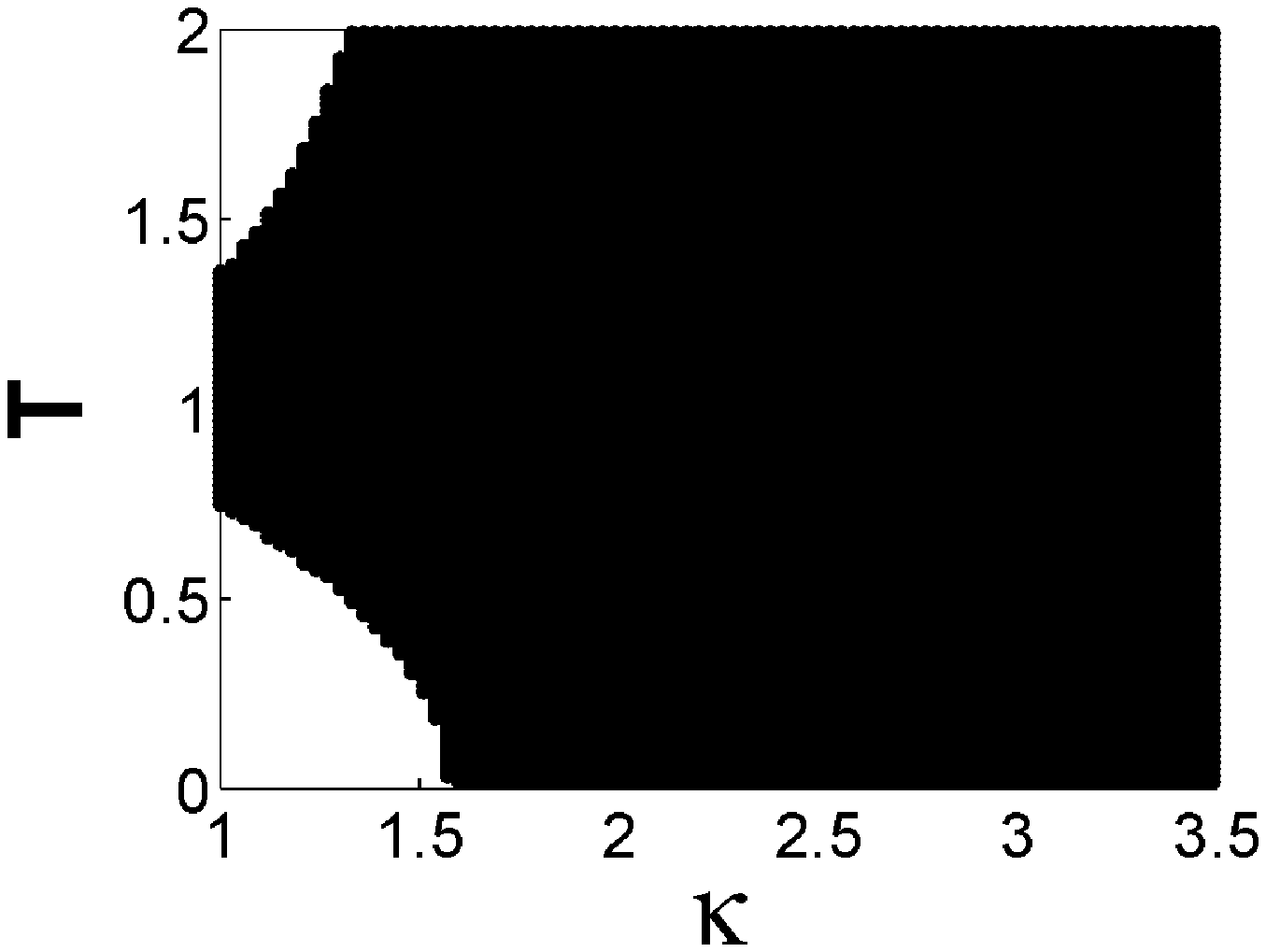}(b)
\renewcommand{\figurename}{Fig.}

\parbox[t]{12.6cm}{\small  Fig.3.  The linear stability parameters regions (white for
stable region and black for unstable region) when $(\kappa,T)$ is in
$(0,13)\times(0,2)$ (a) and in $(1,3.5)\times(0,2)$ (b),
respectively. }\label{fig3}
\end{center}
\end{figure}

\newpage


\begin{thebibliography}{**}

{\small \renewcommand\baselinestretch{.6}


\bibitem{PN}
N. Papanicolaou, J. Phys. A: Math. Gen. {\bf 20} (1987) 3637.

\bibitem{ML}
M. Laksmanan, Phys. Lett. A {\bf 61} (1977) 53.

\bibitem{Ta}
L.A. Takhtajan, Phys. Lett. A {\bf 64} (1977) 235.


\bibitem{Ish} Y. Ishimori, J. Phys. Soc. Japan, {\bf 52} (1982) 3417.


\bibitem{PDL}
K. Porsezian, M. Daniel and M. Lakshmanan, J. Math. Phys. {\bf 33}
(1992) 1907.

\bibitem{DA}
M. Daniel and R. Amuda, Phys. Rev. B {\bf 53} (1996) R2930.

\bibitem{dq4}
Q. Ding,  Phys. Lett. A {\bf 266} (2000), 146.

\bibitem{MS}
H.J. Mikeska and M. Steiner, Adv. Phys. {\bf 44} (1991) 191.


\bibitem{ZT} V.E. Zakharov and L.A. Takhtajan, Theor. Math. Phys.
{\bf 38} (1979) 17.


\bibitem{FT}
L.D. Faddeev and L.A. Takhtajan, Hamiltonian Methods in the Theorey
of Solitons, Springer-Verlag, Berlin, Heideberg 1987.

\bibitem{dq5}
Q. Ding and Z. Zhu, J. Phys. Soc. Japan {\bf 72} No.1 (2003), 49.

\bibitem{dq6}
Q. Ding, J.Phys. A: Math. Theor. {\bf 40} (2007) 1991.

\bibitem{DLS}
F. Delyon, Y.E. Levy and B. Souillard, Phys. Rev. Lett. {\bf 57}
(1986) 2010.

\bibitem{HA} B.M. Herbst, and M.J. Ablowitz, Phys. Rev. Lett. {\bf
62}  (1989) 2065.

\bibitem{WS}
Yi Wan and C.M. Soukoulis, Phys. Rev. A {\bf 41},800 (1990)

\bibitem{HSGT}
D. Hennig, N.G. Sun, H. Gabriel and G.P. Tsironis, Phys. Rev. E {\bf
52} (1995) 255.

\bibitem{Fl}
C. Flytzanis, in {\it Nonlinear Phenomenon in Solids}, edited by
A.F. Vavrek (World Scientific, Singapore, 1985).

\bibitem{KRB}
P.G. Kevrekidis, K. Rasmussen and A.R. Bishop, Inter. J. Mod. Phys.
B {\bf 15} (2001) 2833.

\bibitem{Robin}
C. Robinson, ``Dynamical Systems: Stability, Symbolic Dynamics, and
Chaos," 2nd Ed., CRC Press, 1998.

\bibitem{Grimshaw}
R. Grimshaw, ``Nonlinear Ordinary Differential Equations,"
Blackwell, Oxford, 1990.

}


\newpage
\end{thebibliography}
\end{document}